%% file: main-arxiv.tex
\documentclass[12pt,nofootinbib,notitlepage,preprint,aps,pra,
  amsmath,amssymb,longbibliography]{revtex4-2}
\include{figures}

\include{packages}

\begin{document}
\singlespacing
\renewcommand{\abstractname}{}
\title{Diffusive Braking of Penetrative Convection in Stably-Stratified Fluids}

\author{Bradley W. Hindman$^{1,2,3}$}
\author{J. R. Fuentes$^{4}$}

\affiliation{\small$^{1}$Department of Applied Mathematics, University of Colorado Boulder, Boulder, CO 80309-0526, USA\\$^{2}$JILA, University of Colorado Boulder, Boulder, CO 80309-0440, USA\\$^{3}$Department of Astrophysical and Planetary Sciences, University of Colorado, Boulder, CO 80309, USA\\$^{4}$TAPIR, California Institute of Technology, Pasadena, CA 91125, USA}

\begin{abstract}
Mixing at the interface between a convection zone and an adjacent, stably-stratified layer plays a crucial role in shaping the structure and evolution of stars and planets. In this work, we present a suite of 2D and 3D Boussinesq simulations that explore how bottom-driven convection penetrates into a compositionally stratified region. Our results reveal two distinct regimes: a penetrative regime, where the convection zone steadily grows by entraining fluid from above, and a stalled regime, where growth halts and transitions to overshooting convection. We extend classical entrainment theory by incorporating thermal and compositional diffusion and by deriving a modified entrainment law that predicts interface speeds in the weak-diffusion limit. We show that convection stalls when the interface speed becomes comparable to the compositional diffusion speed and validate the transition between behaviors across a wide parameter space of Richardson and Lewis numbers. Such diffusively-controlled stalling is unlikely to occur in stellar and planetary interiors, where the Lewis number is typically large and compositional diffusion is extremely slow. In these environments, compositional diffusion will merely slow the growth of the convection zone and convective boundaries are expected to stall only in the presence of other curtailing mechanisms such as strong radiative diffusion or rapid rotation.
\end{abstract}

\maketitle

%%%%%%%%%%%%%%%%%%%%%%%%%%%%%%%%%%%%%%%%%%%%%%%%%%%%%%%%%%%%%%%%%%%%%%%%%
%   I. Introduction							                            %
%%%%%%%%%%%%%%%%%%%%%%%%%%%%%%%%%%%%%%%%%%%%%%%%%%%%%%%%%%%%%%%%%%%%%%%%%
{\centering\section{Introduction} \label{sec:Introduction}}

In astrophysics and geophysics, convective layers often sit directly beneath or above stably-stratified regions. When the convective motions are sufficiently fast, convective plumes can overshoot the boundary separating the convection zone from the stable layer, thereby carrying heat and material across the boundary and potentially causing mixing between the layers. If this mixing results in the slow ingestion of the stable layer into the convection zone, this entrainment process is known as penetrative convection \cite[see, e.g.,][]{Anders2022a}. Penetrative convection can play a pivotal role in controlling the dynamics of planetary atmospheres and interiors. For example, in the interior of Earth-like planets and in the gas giants, convection is responsible for generating magnetic fields through a dynamo mechanism \citep[e.g.,][]{Stevenson2003} and penetrative convection is critical in determining the spatial extent of the dynamo region. Another example is in the context of ``fuzzy'' cores in Jupiter and Saturn \cite[e.g.,][]{Howard_2023,Helled2024}, where penetrative convection plays a key role in transporting heavy elements from the core into the hydrogen-rich envelope. In hot Jupiters, where intense stellar radiation creates dynamic atmospheric layers, penetrative convection may contribute to atmospheric mixing and heat redistribution \citep{Showman2020}. Beyond the planetary context, penetrative convection is an important mechanism in stellar astrophysics. The convective cores of massive stars (Type O--F) incinerate their nuclear fuel so rapidly, that continued burning requires transport of fresh fuel from the overlying stable envelope. Hence, penetrative convection regulates the star's luminosity and lifetime by controlling the rate at which fuel is injected into the nuclear furnace \citep[e.g.,][]{Anders2022}.

Of significant interest in many studies of penetrative convection is the speed at which the convective boundary advances through the region of stable stratification, as this speed governs the transport of material into the convection zone. In this context, the response of a compositionally stratified layer of fluid to heating from below (or cooling from above) has been widely studied in laboratory experiments with saltwater \citep[e.g.,][]{Turner1968,Fernando1987} and with numerical simulations \citep[e.g.,][]{Molemaker1997, Fuentes2020, Fuentes2023}. These studies have shown that a well-mixed convective layer develops near the boundary where the heating (or cooling) flux is applied and when the initial stratification has a uniform density gradient, the thickness of this layer grows with time as $h \propto t^{1/2}$.

In such systems, diffusion has generally been relatively unimportant, and the growth of the convection zone has therefore been ubiquitous and inevitable. However, diffusion should not be ignored, as it can either accelerate or delay the motion of the convective boundary depending on the configuration of the laboratory or numerical experiment. In theoretical and numerical studies, weak {\it thermal diffusion} has been shown to slow the speed of the boundary. While the power-law behavior, $h\propto t^{1/2}$, remains unmodified, the prefactor can be significantly reduced \citep{Fuentes2020}. A recent laboratory experiment has suggested that {\it chemical diffusion} may lead to the opposite effect. \citet{Dorel2023} conducted an experimental study of penetrative convection within ambient air laced with a gradient of sulfur hexafluoride gas. Unlike previous work, the initial density profile was Gaussian rather than linear and the effects of chemical diffusion were not negligible. \citet{Dorel2023} found that the convection zone in their experiments grew at a faster rate, $h \propto t^2$, than the classic prediction and attribute this enhanced interface speed to the chemical diffusion of the sulfur hexafluoride.

Conversely, in stellar astrophysics, convection zone boundaries are almost always thought of as static structures on evolutionary timescales.  This is attributed to the large radial gradient in the opacity leading to rapid radial variation of the thermal diffusivity (conduction by photon transport). The convection zone initially grows until thermal conduction becomes sufficiently efficient to suppress convection beyond a certain radius. Once the convective interface reaches this height, the boundary stops moving.  Mixing can still occur across the boundary through convective overshooting, but the boundary itself becomes stationary.  Obviously such stalling behavior is diffusively controlled \citep[e.g.,][]{Andrassy2020,AndersNAT2023}.  %{\bf Do we want to reference anyone in this paragraph?  Andrassi?}

There are several other factors that can change the growth rate of the convective layer. For example, rotation and strong stratification have been shown to decrease the speed of the convective interface. For the case of strong stratification, the slowing of the interface is a simple consequence of energetics; more potential energy is required to mix an atmosphere with strong stability. This stratification effect appears through the buoyancy frequency $N$ of the stable layer, i.e., $h \propto N^{-1} t^{1/2}$ \cite[e.g.,][]{Turner_1965, Turner1968, Fuentes2022}. Rotation on the other hand, instead of changing the energy required to mix, changes the amount of energy that is available to perform lifting work. In a rapidly rotating system, with a rotation rate of $\Omega$, the kinetic energy in the convection is significantly reduced, changing the scaling to $h \propto N^{-5/6} \,\Omega^{-1/4}\,t^{5/12}$ \citep{Fuentes2023, Hindman2023}. While, the rotation and stability of an atmosphere cannot directly cause the convective interface to stop moving (or to stall), we speculate that either effect could indirectly lead to a stall by slowing the interface sufficiently that diffusive effects become important.

Here we will seek to explore in detail how such diffusively controlled stalling of convective penetration occurs, under what conditions it should be expected, and the compositional and thermal profiles that result. We will accomplish this goal by generating a suite of numerical simulations for which we vary both the stability of the fluid layer and the compositional diffusivity. By increasing either, we expect to weaken the growth rate of the thickness of the convective layer and eventually, at extreme enough values, stop the growth altogether. We present analytic solutions for the propagation of convective boundaries in Section~\ref{sec:model}, discuss 2D and 3D numerical simulations of the same processes in Section~\ref{sec:Numerical_Simulations}, and compare the analytic predictions with the simulation results in Section~\ref{sec:Evolution}. We conclude with a summary and a discussion of the implications of our findings in Section~\ref{sec:discussion}.\\

%%%%%%%%%%%%%%%%%%%%%%%%%%%%%%%%%%%%%%%%%%%%%%%%%%%%%%%%%%%%%%%%%%%%%%%%%
%   II. Model of the Entrainment Process					            %
%%%%%%%%%%%%%%%%%%%%%%%%%%%%%%%%%%%%%%%%%%%%%%%%%%%%%%%%%%%%%%%%%%%%%%%%%
{\centering\section{Model of the Entrainment Process}
\label{sec:model}}
When the convective boundary is thin and diffusion is neglected, traditional energetic arguments can be used to predict the speed at which the interface advances.  For the sake of specificity, we assume that the initially stable fluid layer is heated from below, such that the convection zone forms at the bottom and eats upward into the stable layer. In a Boussinesq fluid, this set up is completely equivalent to cooling the fluid from above with the convection zone expanding downwards. In a stratified fluid, as the convection zone expands, a density jump naturally develops at the interface with the stable region.  This density jump forms a gravitational potential energy barrier that opposes further entrainment. The entrainment hypothesis \citep[e.g.,][]{Linden1975} states that the energy to overcome this barrier comes from the kinetic energy of the convective motions; explicitly, the rate of potential energy increase is proportional to the vertical kinetic energy flux carried by the convection. Such a hypothesis leads to a differential equation for the depth of the convection zone, $h$, i.e.,
\begin{equation} \label{eqn:hypothesis}
    \frac{1}{2}g h \, \Delta\rho \frac{dh}{dt} = \frac{\gamma}{2}\rho_0 U^3_{c}\,,
\end{equation}
where $g$ is the gravitational acceleration, $\rho_0$ is the mass density within the convection zone, $\Delta\rho$ is the density jump across the thin interface, $U_c$ is a typical convective flow speed, and $\gamma$ is a dimensionless constant of proportionality called the mixing efficiency. The left-hand side of Equation~(\ref{eqn:hypothesis}) is the rate of the change of the potential energy. This formulation is correct for any initial density stratification as long as the mass density in the convection zone is well mixed.

This entrainment equation can be solved for the thickness of the convection zone $h$ and the speed of the convective boundary $dh/dt$ given prescriptions for the density jump across the convective interface $\Delta\rho$ and the convective speed $U_c$. Typically, the density jump is specified by assuming that the convection zone is well-mixed and the stable layer remains unmodified until the fluid is entrained into the convection zone (i.e., diffusive restratification is ignored). Under such assumptions, the density jump can be expressed in terms of the buoyancy frequency of the initial stable stratification, $N_0$.  For a spatially uniform buoyancy frequency, the density jump is proportional to the depth of the convection zone,

\begin{equation}
    \label{eqn:mixing}
    \Delta\rho = \frac{\rho_0 N_0^2 h}{2g} \,.
\end{equation}

\noindent For the convective velocity, mixing length theory \cite[e.g.,][]{Fuentes2020} is usually employed to provide an estimate. Without rotation, this estimate becomes
\begin{equation}
    \label{eqn:Uc}
    U_c = \left(\frac{g\alpha F_*}{\rho_0 c_p} \, h\right)^{1/3} \; ,
\end{equation}
\noindent where $\alpha$ is the coefficient of thermal expansion, $c_p$ is the specific heat capacity at constant pressure, and $F_*$ is the heat flux forced through the convection zone.

Putting these ingredients together and assuming that the heat flux and the mixing efficiency are constants generates the classic entrainment law for the depth of the convection zone,

\begin{equation}
    \label{eqn:classic_h(t)}
    h(t) = 2\left(\gamma+ \frac{1}{2}\right)^{1/2} \left(\frac{g \alpha F_* t}{\rho_0 c_p N_0^2}\right)^{1/2} \propto F_*^{1/2} N_0^{-1} t^{1/2}\,.
\end{equation}
As expected, the convection zone propagates more rapidly with a larger imposed heat flux and more slowly with a steeper initial density gradient, i.e., larger buoyancy frequency.
\\

%===============================================================%
%   IIA. Diffusive Effects                                      %
%===============================================================%
{\centering\subsection{Diffusive Effects}}

A primary weakness of Equation~(\ref{eqn:classic_h(t)}) is the lack of diffusive effects in Equations~(\ref{eqn:hypothesis}) and (\ref{eqn:mixing}). While chemical diffusion can often be ignored under many astrophysical and geophysical conditions, if the convective boundary were to ever stall (e.g., as has been suggested for Jupiter's convection zone \cite{Hindman2023}), diffusion is likely to adjust the quasi-equilibrium that would be established. To address this weakness, we follow the energetic arguments first established by \citet{Linden1975}, and subsequently expanded by \citet{Fernando1987} and \citet{Fuentes2020}. We adapt these previous derivations of the speed of the convective boundary by considering the effects of weak chemical diffusion of a heavy solute. For simplicity we work in a Boussinesq framework \citep[e.g.,][]{Spiegel_Veronis_1960} where the mass density, $\rho = \rho_0 (1 - \alpha T + \beta C)$, can be expressed as a linear combination of the potential temperature $T$ and the solute concentration $C$.  Here, $\rho_0$ is a fiducial density and $\alpha$ and $\beta$, are the coefficients of thermal expansion and compositional contraction (both assumed to be positive constants).

Since, mass can be redistributed both by mixing and chemical diffusion, Equation~(\ref{eqn:hypothesis}) must be augmented when diffusion is considered, 
\begin{equation}
    \frac{1}{2} gh \left(\Delta\rho \frac{dh}{dt} + \rho_0 \beta \kappa_C \left|\partial_z C\right|_\cb\right) = \frac{\gamma}{2} \rho_0 U_c^3 \; .
    \label{eqn:dEdt}
\end{equation}
\noindent As in Equation~(\ref{eqn:hypothesis}), the first term within the parentheses represents the rate of work done to mix the light fluid from the stable layer into the convection zone. The additional diffusive term represents the energy required to keep the convection zone well-mixed as diffusion removes solute from the upper portion of the convection zone. This energy tax depends directly on the compositional diffusive flux which depends on the chemical diffusivity for the heavy solute, $\kappa_C$, and the concentration gradient at the base of the stably stratified envelope, just above the convective boundary, $\left|\partial_z C\right|_\cb$.

The jump in the mass density across the convective boundary is related to the corresponding jumps in the composition and potential temperature, $\Delta \rho = \rho_0(\beta \Delta C - \alpha \Delta T)$.  An estimate for each of these jumps can be obtained by considering conservation of solute and by noting that the convection zone warms with time due to the heat flux from below, $F_*$, and cools due to the conductive heat flux $F_{\rm cb}$ upwards across the convective boundary,
\begin{equation}
    \Delta\rho = \frac{\rho_0 N_0^2h}{2g} - \left(\alpha \frac{F_* - F_\cb}{c_p} + \rho_0\beta\kappa_C |\partial_z C|_\cb \right) \frac{t}{h} \; .
    \label{eqn:deltarho}
\end{equation}

\noindent The conductive flux depends on the thermal diffusivity $\kappa_T$ and the temperature gradient $|\partial_zT|_\cb$ just above the convective boundary, $F_{\rm cb}=\rho_0 c_p \kappa_T |\partial_z T|_\cb$. The first term on the right-hand side of Equation~(\ref{eqn:deltarho}) is identical to that appearing in Equation~(\ref{eqn:mixing}) and arises from mixing of the initial density profile and thus involves the buoyancy frequency of the initial stratification, $N_0 \equiv \sqrt{g\left(\alpha\partial_z T_0 - \beta\partial_z C_0\right)}$. For simplicity and specificity, we have assumed that the initial stratification has a uniform buoyancy frequency. The second term represents changes in the convection zone's mass density due to thermal and compositional diffusion at its boundaries. We assume that $F_\cb$ and $\left|\partial_z C\right|_\cb$ are both temporally constant such that the diffusive contribution to the density jump---the second term on the right-hand side of Equation~(\ref{eqn:deltarho})---can be written as a linear function of time. This assumption has been shown to be valid when diffusion leads to only weak restratification of the stable fluid above the convective interface \citep{Fuentes2020}.

As in previous studies, we use mixing-length theory to provide an estimate for the convective speed, Equation~(\ref{eqn:Uc}). The reader should note, however, that another interpretation of Equation~(\ref{eqn:Uc}) is that the ultimate source of energy required for mixing is the gravitational potential energy pumped through the lower boundary by the heat flux. The injection rate is given by $g\alpha F_* h /c_p$, which is the quantity that appears on the right-hand side of Equation~(\ref{eqn:Uc}).

Combine Equations~(\ref{eqn:Uc})--(\ref{eqn:deltarho}) to obtain the following nonlinear ODE for the depth of the convection zone $h$ as a function of time,
\begin{equation}
    \left[1 - \left(\frac{1-\varepsilon_T}{\Ri} + \frac{\varepsilon_C}{\Le}\right) \frac{2\kappa_T t}{h^2}\right] \frac{h}{2\kappa_T} \frac{dh}{dt}  = \frac{\gamma}{\Ri} - \frac{\varepsilon_C}{\Le}  \; .
    \label{eqn:ODE}
\end{equation}
\noindent where Le is the Lewis number and Ri is a global Richardson number,
\begin{equation}
    \Le \equiv \frac{\kappa_T}{\kappa_C}\,,  \qquad \Ri \equiv \frac{\rho_0 c_p\kappa_T N_0^2}{g\alpha F_*}~. %= \frac{H^2 N_0^2}{U_{\rm ff}^2}\,.
\end{equation}
The  Richardson number encodes the stability of the fluid layer relative to the thermal driving of convective overturning.  High Richardson numbers indicate strong stratification with a high level of stability. Finally, in Equation~(\ref{eqn:ODE}), the diffusive fluxes at the base of the stable envelope have been nondimensionalized,
\begin{equation}
    \varepsilon_T \equiv \frac{F_\cb}{F_*} \,,   \qquad
    \varepsilon_C \equiv \frac{\beta|\partial_z C|_\cb}{(N_0^2/g)} \,.
\end{equation} 

Equation~(\ref{eqn:ODE}) has a general solution for time $t$ as a function of convection zone thickness $h$,
\begin{eqnarray}
    t(h) &=& t_0 \left(\frac{h}{H}\right)^{-a} + \frac{h^2}{4\kappa_T V_{\rm cb}^2} \; ,
\\
    V_{\rm cb}^2 &\equiv& \left(\gamma + \frac{1-\varepsilon_T}{2}\right) \Ri^{-1} - \frac{\varepsilon_C}{2} \Le^{-1} \; ,
    \label{eqn:Vi}
\\
    a &\equiv& \frac{(1-\varepsilon_T) \Ri^{-1} + \varepsilon_C \Le^{-1}}{\gamma \Ri^{-1} - \varepsilon_C \Le^{-1}} \; ,
\end{eqnarray}

\noindent where $t_0$ is an arbitrary constant (used to satisfy the initial condition) and $V_{\rm cb}$ is a dimensionless constant that characterizes the speed of the convective boundary. The homogeneous solution, $(h/H)^{-a}$, is a transient that decays away as the convection zone deepens. The particular solution, $h^2/(4\kappa_T V_{\rm cb}^2)$, generates the classic $h \propto t^{1/2}$ power-law solution, 
\begin{equation}
    \label{eqn:hcz} 
    h(t) = 2 V_{\rm cb} \, \left(\kappa_T t\right)^{1/2} , \quad
    \frac{dh}{dt} = V_{\rm cb} \, \left(\frac{\kappa_T}{t}\right)^{1/2} ,
\end{equation}

\noindent In the absence of thermal and compositional diffusion ($\varepsilon_T \to 0$ and $\Le^{-1} \to 0$), the convection zone's depth reduces to the traditional entrainment law, Equation~(\ref{eqn:classic_h(t)}). The presence of weak thermal or compositional diffusion slows the progression of the convective boundary (through $V_{\rm cb}$) but does not change the power-law relationship with time.
\\

%===============================================================%
%   IIB. Characteristic Propagation Speeds                      %
%===============================================================%
{\centering\subsection{Characteristic Propagation Speeds}}

The speed of the convective interface, Equation~(\ref{eqn:hcz}), has been derived under the assumption that diffusion acts slowly enough that diffusive restratification of the stable region above the convective interface can be ignored.  Further, we have also assumed that the convective interface is thin so that the density jump can be treated as a discontinuity. Both of these approximations are justified if the speed of diffusive spread is slow compared to the speed of the interface.  Under such conditions, the diffusive widening of the convective boundary is balanced by the sharpening of the interface caused by entrainment.

A diffusive boundary layer widens self-similarly with a speed that is a temporal power law, $V = (\kappa/t)^{1/2}$, where $\kappa$ is the relevant diffusivity. Conveniently, this is the same power law as derived for the speed of the entrainment front, such that we can easily compare diffusive speeds with the interface speed. For example, our assumptions should be valid as long as the compositional diffusion speed, $V_C = (\kappa_C/t)^{1/2}$, is slower than the speed of the convective interface,

\begin{equation} \label{eqn:Vcb=VC}
   \left(\frac{\kappa_C}{t}\right)^{1/2} < V_\cb\left(\frac{\kappa_T}{t}\right)^{1/2} \Longrightarrow \Le^{-1/2} < V_\cb  \Longrightarrow \Ri < \frac{2\gamma+1-\varepsilon_T}{1+2\varepsilon_C} \Le\,.
\end{equation} 

\noindent  The final inequality above is obtained through use of Equation~(\ref{eqn:Vi}).

When compositional diffusion is large, such that the previous condition is not satisfied, we expect the convective boundary to diffusively widen and the stable layer to be restratified above the interface.  This has the potential to weaken the buoyancy frequency and lead to faster entrainment (this is the argument made by \citet{Dorel2023} to explain the rapid advance of the convective boundary in their laboratory experiment).  However, once the stable layer starts to restratify, we need to consider the boundary conditions on the upper surface. We will find that in our suite of simulations, our choice of boundary conditions permits a quasi-equilibrium to be achieved after a compositional diffusion time passes. This equilibrium is one in which the convection zone is no longer growing, the density jump has disappeared, and the energy available for mixing is completely consumed by mixing the mass deficit arising from compositional diffusion out of the convection zone.
\\

%%%%%%%%%%%%%%%%%%%%%%%%%%%%%%%%%%%%%%%%%%%%%%%%%%%%%%%%%%%%%%%%%%%%%%%%%
%   III. Numerical Simulations          					            %
%%%%%%%%%%%%%%%%%%%%%%%%%%%%%%%%%%%%%%%%%%%%%%%%%%%%%%%%%%%%%%%%%%%%%%%%%
{\centering\section{Numerical Simulations}
\label{sec:Numerical_Simulations}}
In order to test the validity of Equation~(\ref{eqn:hcz}), we present a suite of 2D and 3D simulations performed in Cartesian boxes where convection is driven from below and the ensuing turbulence chews its way upward through a stable region with uniform buoyancy frequency. We systematically vary the diffusivities and the buoyancy frequency of the initial stratification by varying the Lewis and Richardson numbers. In particular, we are curious if the advance of the convection boundary can stall because a diffusive quasi-equilibrium is achieved. The 2D simulations are used to examine a broad parameter space, whereas the more limited sampling of 3D simulations are used to test the validity of the 2D results. In all simulations, the height of the Cartesian box is $H$ with gravity antiparallel to the $z$-coordinate axis, $\bm{g} = -g \hat{\bm{z}}$, and the bottom of the box corresponds to $z=0$. The 2D simulations have a horizontal width of $2H$ to avoid the onset of artificial shear flows \citep{Fuentes2021}, whereas the 3D simulations have a width of $H$ in both horizontal directions.

Since the stratification affects the rate of entrainment primarily through the buoyancy frequency, for simplicity, we choose an initially uniform potential temperature, setting $T_0(z)=0$, and capture the atmosphere's stability with a linear concentration profile, $C_0(z) = (\beta g)^{-1} N_0^2 (H - z)$. While the vertical profile of the initial mass density is stable, convection will be driven by applying a potential temperature gradient as a boundary condition on the lower boundary of the domain. We express this gradient in terms of the heat flux, $F_*$, passing through the lower surface, $\partial_z T|_{z=0} = -F_*/(\rho_0 c_p \kappa_T)$.\\

%===============================================================%
%   IIIA. Nondimensional Boussinesq Fluid Equations             %
%===============================================================%
{\centering\subsection{Nondimensional Boussinesq Fluid Equations}}
We numerically solve the fluid equations in nondimensional form. We adopt the layer height $H$ as the characteristic length scale, the thermal diffusion time $\tau_T = H^2/\kappa_T$ as the unit of time, $[T] = H|\partial_z T|_{z=0}$ as the unit of temperature, and $[C] = \alpha [T] / \beta$ as the unit of concentration.  With this choice, the Boussinesq fluid equations take on the following nondimensional form,

\begin{eqnarray}
    \frac{\partial \bm{u}}{\partial t} + \bm{u}\cdot\grad\bm{u} &=& -\grad \varpi - \Ra\Pr\, (C-T) \, \hat{\bm{z}} + \Pr\nabla^2 \bm{u}\,,
    \label{eqn:Navier-Stokes}
\\
    \frac{\partial T}{\partial t} + \bm{u}\cdot\grad T &=& \nabla^2 T\,,
\\
    \frac{\partial C}{\partial t} + \bm{u}\cdot\grad C &=& \Le^{-1}\nabla^2 C\,,
\\
    \grad\cdot\bm{u} &=& 0 \, .
    \label{eqn:Incompressibility}
\end{eqnarray}

\noindent In these equations, $\bm{u}$ is the nondimensional velocity and $\varpi$ is the nondimensional reduced pressure. In the momentum equation, hydrostatic balance involving the fiducial density has been removed. All other variables have their previous definitions (e.g., $T$ is potential temperature) except they are now nondimensional. 

There are four dimensionless numbers that characterize the evolution of the flow. In addition to the global Richardson number and the Lewis number that we have already introduced, the flux Rayleigh number $\Ra$ and the Prandtl number are important,
\begin{eqnarray}
    \Ra &\equiv& \frac{g \alpha F_* H^4}{\rho_0 c_P \kappa_T^2 \nu}  \, , \quad\;
    \Pr \equiv  \frac{\nu}{\kappa_T}%\Ri \equiv \frac{\rho_0 c_p \kappa_T N_0^2}{g \alpha F_*} =  \frac{N_0^2 H^2}{U_{\rm ff}^2} \,,
%\\
    %\Pr &\equiv&  \frac{\nu}{\kappa_T} %\,, \qquad\qquad %\Le \equiv \frac{\kappa_T}{\kappa_C} \, ,
\end{eqnarray}

\noindent where $\nu$ is the kinematic viscosity. The Rayleigh, Lewis, and Prandtl numbers appear directly in the nondimensional fluid equations. The Richardson number appears through the initial and boundary conditions.  Our initial conditions are a static diffusive equilibrium, i.e.,

\begin{equation}
    \bm{u}(t=0) = 0\, \qquad T(t=0) = T_0(z) = 0\,, \qquad C(t=0) = C_0(z) =  \Ri (1-z) \,.
\end{equation}

\noindent We enforce periodic horizontal boundary conditions on all fluid variables. The upper and lower boundaries are impenetrable and stress-free,

\begin{equation}
    \bm{u}\cdot\hat{\bm{z}} = 0\, \qquad \partial_z \bm{u}\times \hat{\bm{z}} = 0\,.    
\end{equation}
\noindent We apply fixed-flux boundary conditions on the potential temperature and concentration,
\begin{equation}
    \partial_z T(z=0) = -1\,, \qquad \partial_z T(z=1) = 0\,, \qquad \partial_z C(z=0) = 0\,, \qquad \partial_z C(z=1) = -\Ri \;.    
\end{equation}

\noindent A schematic of our numerical setup is presented in Figure~\ref{fig:scheme}.

\figschematic

In many astrophysical and geophysical contexts, all of these nondimensional numbers take on extreme values. For example in stars, the Rayleigh number and Lewis number are enormous, $\Ra\sim10^{24}$ and $\Le \sim 10^6$, while the Prandtl number is tiny, $\Pr \sim 10^{-6}$ \citep{Jermyn2022}. While the Rayleigh number is similar in gas giants, the Prandtl and Lewis number for Jovian planets are less extreme, with $\mathrm{Pr}\sim 10^{-2}$ and $\mathrm{Le}\sim 10^{2}$ \citep[][]{French2012}. Also, from the estimates given in \cite{Jermyn2022} and \cite{Lecoanet2023}, the radiative zone of a solar-like star has $N\sim  10^{-3}~{\rm{s^{-1}}}$, a photon thermal diffusivity $\kappa_T \sim 10^9~\mathrm{cm^2~s^{-1}}$, $\rho_0 \sim 1~\mathrm{g~cm^{-3}}$, $g\sim 10^{4}~\mathrm{cm~s^{-2}}$, $c_p\sim 10^8~\mathrm{erg~g^{-1}~K^{-1}}$, $\alpha \sim 10^{-6}~\mathrm{K^{-1}}$, and $F_* \sim 10^{10}~\mathrm{erg~cm^{-2}~s^{-1}}$. These values yield a global Richardson number of order $\mathrm{Ri}\sim 10^{3}$, although we note that the precise value can vary significantly among stellar models. On the other hand, in the compositional fuzzy cores of gas giants, $N\sim 10^{-4}~{\rm{s^{-1}}}$, $\rho_0 \sim 3~\mathrm{g~cm^{-3}}$, $\kappa_T \sim 10^{-1}~\mathrm{cm^2~s^{-1}}$, $F_*\sim 5\times 10^3~\mathrm{erg~cm^{-2}~s^{-1}}$, $g\sim 2\times 10^3~\mathrm{cm~s^{-2}}$, $c_p \sim 2\times 10^8~\mathrm{erg~g^{-1}~K^{-1}}$, and $\alpha\sim 10^{-5}~\mathrm{K^{-1}}$ \citep{Stevenson1977,Guillot2004}. Unlike in stars, these values yield $\mathrm{Ri} \sim 10^{-2}$ for giant planets. Realistic astrophysical and geophysical values of the Rayleigh, Lewis, Prandtl, and Richardson numbers are inaccessible in current simulations, so the ones implemented in our simulations are more moderate, but the important issue is that we have captured the correct ordering of time scales $\tau_{\rm ff} < \tau_T < \tau_\nu \approx \tau_C$, where $\tau_{\rm ff} = H/U_{\rm ff}$ is the freefall time, $\tau_T = H^2/\kappa_T$ is the thermal diffusion time, $\tau_\nu = H^2/\nu$ is the viscous diffusion time, and $\tau_C = H^2/\kappa_C$ is the compositional diffusion time.

Our suite of 2D simulations explores a grid of models in Ri--Le space, each number taking on one of five different possibilities, $\Ri \in [1, 3, 5, 7, 10]$ and $\Le \in [1, 2, 4, 6, 10]$. The Rayleigh number and Prandtl number were kept constant ($\Ra = 10^8$ and $\Pr = 0.1$). Figure~\ref{fig:regimes} graphically presents the grid of 2D simulations in Ri--Le space. The 3D simulations were fewer in number, varying the Richardson number over three values $\Ri \in [1,5,10]$ while keeping the Rayleigh, Prandtl and Lewis numbers fixed ($\Ra= 10^8$, $\Pr=0.1$, and $\Le = 10$).\\

\figregimes

%===============================================================%
%   IIIB. Numerical methods                                     %
%===============================================================%
{\centering\subsection{Numerical methods}
\label{subsec:numerical_methods}}

We evolve Equations (\ref{eqn:Navier-Stokes})--(\ref{eqn:Incompressibility}) using version 3 of the Dedalus pseudospectral solver \citep{Burns2020}, with a Runge-Kutta timestepper RK443 (see \S2.8 of \cite{Ascher1997}) and a Courant–Friedrichs–Lewy (CFL) safety factor 0.25. In the 2D simulations, all variables are represented using a spectral expansion composed of 512 Chebyshev polynomials in the vertical coordinate and 1024 Fourier modes in the horizontal coordinate. In 3D, we employ 384 Chebyshev polynomials and 384 Fourier modes in both horizontal directions. We use the 3/2-dealiasing rule in all directions, such that nonlinearities are calculated in physical space with a $1536 \times 768$ mesh in 2D or a $576^3$ grid in 3D.  We initialize the simulations with random noise in the temperature field, with each value drawn from a normal distribution with a magnitude of $10^{-5}$.\\

%%%%%%%%%%%%%%%%%%%%%%%%%%%%%%%%%%%%%%%%%%%%%%%%%%%%%%%%%%%%%%%%%%%%%%%%%
%   IV. Evolution of the Concentration and Temperature                  %
%%%%%%%%%%%%%%%%%%%%%%%%%%%%%%%%%%%%%%%%%%%%%%%%%%%%%%%%%%%%%%%%%%%%%%%%%
{\centering\section{Evolution of the Concentration and Temperature}
\label{sec:Evolution}}

The initial conditions have a stable stratification (i.e., $N_0^2 > 0$ for all $z$) and are in a diffusive equilibrium (spatially constant composition and heat fluxes). However, the boundary conditions that we apply at the lower boundary (e.g., nonzero heat flux and zero composition flux) are inconsistent with the initial conditions. Hence, immediately after the simulation starts, diffusive boundary layers form on the lower surface in both the temperature and concentration. These boundary layers widen with time and eventually the unstable layer of fluid at the base of the domain achieves sufficient thickness that it becomes convectively unstable and hot plumes rise from the lower boundary.  After a few overturnings, the convection zone becomes fully formed and well mixed. The boundary between the convection zone and overlying stable layer begins to move upward and entrainment commences. %At this point in time, the individual members of our suite of simulations diverge in their behavior.

In roughly half of the simulations (shown with the black squares and circles in Figure~\ref{fig:regimes}), the convective interface ascends rapidly and eventually mixes the entire domain. Because the convection zone continually thickens, these simulations are in a regime of ``penetrative convection". The other half of the simulations (indicated with red disks and circles in Figure~\ref{fig:regimes}) have a convective interface that ascends slowly and eventually stalls at a finite depth. Once such a stall has occurred, a quasi-equilibrium is achieved whereby the box continues to warm and become depleted of solute, but the shapes of the concentration and temperature profiles stop evolving. Convective motions continue to overshoot into the stable layer, but the convection zone stops growing in space.

As predicted earlier, the models with a stalled interface are those with the highest compositional diffusion rates (large $\Le$) and strongest stratification (highest Richardson numbers $\Ri$). The boundary in $\Ri$--$\Le$ space between those models with a rapidly penetrating interface and those with a stalled interface is well matched by the line given by $V_\cb = \Le^{-1/2}$, which corresponds to parameter values for which the speed of the convective interface speed equals the compositional diffusion speed, see Equation~(\ref{eqn:Vcb=VC}). In Figure~\ref{fig:regimes}, the dotted blue curve shows this theoretical boundary for parameter values $\gamma = 0.50$, $\varepsilon_T = 0.71$, and $\varepsilon_C = 0.96$. These values were chosen by measuring the thickness of the convection zone $h(t)$ over our suite of models and fitting $h = q^{-1} V_\cb \, (\kappa_T t)^q$ to obtain the speed $V_\cb$ and power-law index $q$ for each simulation. Then, we fit the set of speeds for $\gamma$, $\varepsilon_T$, and $\varepsilon_C$, using Equation~(\ref{eqn:Vi}) for the functional form.  In the fitting function, there is an indeterminacy between the mixing efficiency $\gamma$ and the thermal flux ratio $\varepsilon_T$; therefore, we have fixed $\gamma$ to a value of one-half when performing the fits. This choice is motivated by the thorough analysis of the transport across the interface by \citet{Fuentes2020}, who found that $\gamma \approx 0.5$--1 at low Prandtl numbers.

%===============================================================%
%   IVA. Fast Penetrating Convection                            %
%===============================================================%
{\centering\subsection{Fast Penetrating Convection}
\label{subsubsec:rapid_entrainment}}

A 3D simulation ($\Ri=5.0$, $\Le=10.0$) with a quickly ascending interface is illustrated in Figure~\ref{fig:fast_stills} with three snapshots showing the vertical velocity. Figure~\ref{fig:fast_profs}a shows vertical profiles of the mass density for a 3D penetrative-convection model with a weaker stratification ($\Ri = 1$, $\Le = 10$). These profiles have been obtained by averaging the flow field horizontally and over a short temporal window. As the well-mixed convection zone grows, it mixes the atmosphere's initial concentration profile and a jump in the concentration appears across the convective boundary. Similarly, due to heating at the lower boundary, the convection zone also warms and similar jumps appear in the temperature. These two effects lead to a corresponding jump across the boundary in the mass density. These jumps are clearly visible as inflection points in the vertical profiles appearing in Figure~\ref{fig:fast_profs}a.  Similar behavior is observed in all of the simulations represented by the black squares in Figure~\ref{fig:regimes}. 

\figfaststills
\figfastprofs

In this limit of rapid entrainment, the depth of the convection zone does indeed follow the classic $h\propto t^{1/2}$ entrainment law. Figure~\ref{fig:fast_profs}b shows the height of the convective boundary as a function of time for both the 2D and the 3D simulations. We measure this height by seeking extrema in the second derivative with respect to height of the horizontally averaged concentration profiles. For the 2D model, the measurement of the height is noisier because of averaging only over a single horizontal dimension.  Hence, the temporal averaging has also been performed over a longer window to decrease the convective noise. After an initial transient, the height asymptotes to a power-law with the expected index of 1/2. Once the interface approaches close enough to the upper boundary of the simulation domain that overshooting convective plumes begin to hit that boundary, the smooth ascent of the convective interface is disrupted.  In the figure, we remove large times for which this boundary interaction has obviously occurred.

Figures~\ref{fig:fast_profs}c,d show profiles for the vertical fluxes of heat and solute for a single time frame during the simulation. We present the diffusive flux and the turbulent flux in addition to the total flux. The time at which the fluxes were computed is indicated in Figure~\ref{fig:fast_profs}b with the violet dot. Convection has engulfed the lower half of the domain where heat and composition are carried exclusively by the turbulence.  Whereas, in the relatively untouched region above the convection zone, the fluxes are largely diffusive. We note that the turbulent solute flux within the stable region is small but nonzero in Figure ~\ref{fig:fast_profs}d. The bumps and wiggles oscillate rapidly in time and are the result of gravity waves.  These waves play essentially no role in the energetics of the interface. The convective boundary appears most distinctly in the composition flux, as the region with the large negative gradient. Within the convective boundary, mass is being rapidly deposited as the convection carries solute from below and mixes it with the lighter material in the overlying stable region.\\

%---------------------------------------------------------------%
%   IVA1. Verifying the Entrainment Law                         %
%---------------------------------------------------------------%
{\centering \subsubsection{Verifying the Entrainment Law}}
Equation~(\ref{eqn:hcz}) provides an analytic entrainment law that describes how rapidly the convective boundary ascends when diffusive effects are small. Using our suite of 2D simulations, we test this formula by measuring the speed of the interface and comparing it to the theoretical behavior as a function of Richardson and Lewis number.  Since, we expect Equation~(\ref{eqn:hcz}) to hold only after an initial transient has decayed away, we measure the velocity $V_{\rm cb}$ from each simulation by fitting $h(t)$ only for late times with a power law of the form $h(t) = q^{-1} V_{\rm cb} \, (\kappa_T t)^q$.

\figspeed

Figure~\ref{fig:Vcb} presents the resulting speed measurements as a function of the Richardson and Lewis numbers. Every 2D model undergoing penetrative convection is included as a square symbol. In Figure~\ref{fig:Vcb}a, different Lewis numbers are indicated by the color. The colored dotted curves present the analytic formula from Equation~(\ref{eqn:Vi}). The purple curve indicates a theoretical result for an asymptotically large value of the Lewis number. Figure~\ref{fig:Vcb}b presents the same data, but the color now indicates the Richardson number. For the analytic expressions illustrated in Figure~\ref{fig:Vcb}, we have used $\gamma = 0.5$, $\varepsilon_T = 0.71$, and $\varepsilon_C = 0.96$, the same values used to generate the transition curve in the regime diagram shown in Figure~\ref{fig:regimes}. Further, these values were obtained by fitting the 2D simulation results to Equation~(\ref{eqn:Vi}) for all simulations exhibiting a fast convective interface with $\Ri > 1$.

In general, the simulations with Richardson number greater than 1.0 are well matched by the theory.  However, the simulations with $\Ri = 1.0$ all possess significantly faster interfaces than the theory would suggest.  Several factors are at play in this case.  First, the simulations with $\Ri = 1.0$ have such fast interfaces that they often hit the upper surface of the computational domain before they have fully asymptoted to the expected $h\propto t^{1/2}$ power law. Second, for weak stratification (as holds for $\Ri = 1.0$) overshooting hot plumes can extend a significant distance beyond the convective interface and these plumes interact with the upper boundary, causing additional mixing within the stable layer, a weakening of the stratification, and a corresponding enhancement of the interface speed. In future work, deeper boxes (larger Rayleigh number in our notation) should allow direct testing of the entrainment law for low Richardson numbers.\\

%===============================================================%
%   IVB. Stalled, overshooting convection                       %
%===============================================================%
{\centering\subsection{Stalled, overshooting convection}}

For simulations with strong stratification or high compositional diffusion (see Figure~\ref{fig:regimes}), the convective boundary eventually stalls after a period of initial growth. Warm convective plumes continue to overshoot into the overlying stable stratification, but the convection is no longer penetrative. Figure~\ref{fig:stalled_profs} presents vertical profiles of the mass density and fluxes for our 3D model with $\Ri=10$, $\Le=10$, which has a Richardson number that is ten times higher than the penetrative convection model discussed in the previous subsection. The convective boundary in this model stops ascending at a height of roughly 50\% of the domain height. The stalling of the interface can be clearly seen in Figure~\ref{fig:stalled_profs}b.

\figstalledprofs

In addition to stalling, the profiles for both potential temperature and concentration achieve a quasi-equilibrium for which the vertical mean of each profile continues to evolve linearly with time, but the gradient of the profile becomes steady. Such "equilibria" arise because the fluxes of heat and composition become linear functions of height (see Figures~\ref{fig:stalled_profs}c,d). Since the divergence of each flux is spatially constant, the rate of change of temperature and concentration is spatially uniform. For the nondimensional temperature this balance becomes,

\begin{equation}
    \frac{\partial T}{\partial t} =   -\frac{\partial}{\partial z} \left(\frac{F_{\rm heat}(z)}{F_*}\right) \to -\frac{\partial}{\partial z} \left(1-z\right) = 1 \;, 
\end{equation}

\noindent with the solution

\begin{equation}
    T(z,t) \to t + \theta(z) \,
\end{equation}

\noindent where $\theta(z)$ is a time-independent profile that is self-consistent with the thermal flux. A similar balance is achieved for the composition.

We emphasize that the jumps at the convective boundary that were characteristic of the penetrative convection regime are absent (see Figure~\ref{fig:stalled_profs}a).  Hence, the analytic entrainment law derived in section~\ref{sec:model} is invalid. All of the models indicated by red circles and disks in Fig.~\ref{fig:regimes} display similar stalling behavior, resulting from the evolution toward a diffusive equilibrium. This common behavior is illustrated in Figure~\ref{fig:stall_time} which shows the normalized height of the convective interface as a function of time for all of the models that exhibit a stalled convective boundary.  The height of the interface is normalized by the asymptotic height as time becomes large. Time is shown in compositional diffusion times since the formation of the convection zone. All of the models exhibit a similar temporal behavior and approach the stalled equilibrium after roughly one-half of a compositional diffusion time.\\

\figstalltime

%%%%%%%%%%%%%%%%%%%%%%%%%%%%%%%%%%%%%%%%%%%%%%%%%%%%%%%%%%%%%%%%%%%%%%%%%
%   V. Summary and Discussion                                           %
%%%%%%%%%%%%%%%%%%%%%%%%%%%%%%%%%%%%%%%%%%%%%%%%%%%%%%%%%%%%%%%%%%%%%%%%%
{\centering\section{Summary and Discussion}\label{sec:discussion}}

Through a suite of 2D and 3D simulations of bottom-driven convection, we have found two regimes of behavior. For models with weak stratification (small buoyancy frequency or equivalently small Richardson number $\Ri$) or low compositional diffusion (large Lewis number $\Le$) we find a regime of penetrative convection. The convection zone that develops at the bottom boundary continuously entrains fluid from the overlying stable layer and eventually mixes through the entire computational domain. Conversely, models with strong stratification or high compositional diffusivity reach a quasi-equilibrium for which the convection zone stops growing and settles into a regime of overshooting convection. The boundary between these two regimes appears to coincide with the curve in parameter space where the speed of the convective interface (as expressed analytically in Equation~(\ref{eqn:hcz}) equals the compositional diffusion speed. The fate of the system is essentially the same in 2D and 3D. The main distinction, as expected from the inverse cascade in 2D turbulence, is that the 2D cases mix more efficiently than the 3D ones.

%{\centering\subsection{Diffusive Effects on the Interface Speed}}

We find that diffusion slows the advance of the convective boundary in our analytic model and in all of our simulations. Figure~\ref{fig:Vcb} illustrates how increasing the compositional diffusion rate (or equivalently decreasing the Lewis number) results in a slower interface in the simulations. In our analytic model, the convective boundary ascends with a speed given by

\begin{equation}
    V = \left[\left(\gamma+\frac{1-\varepsilon_T}{2}\right) \frac{\kappa_T}{\Ri} - \frac{\varepsilon_C}{2} \kappa_C\right]^{1/2} t^{-1/2} \;,
\end{equation}

\noindent which is obtained by combining Equations~(\ref{eqn:Vi}) and (\ref{eqn:hcz}). As the diffusivities increase, the interface speed decreases because $\varepsilon_T$ increases and the second term in square brackets increases. The reader should note that $\kappa_T/\Ri$ is independent of the thermal diffusivity; hence thermal diffusion only manifests through the heat flux ratio $\varepsilon_T = F_\cb / F_*$. Physically, all of these diffusive effects manifest through an increase in the density jump across the interface and a reduction of the amount of kinetic energy that is available for mixing the density profile.

In the absence of diffusion, the density jump across the interface is controlled by mixing and heating within the convection zone.  Continual mixing of the initial concentration gradient to a uniform value within the convection zone results in a growing concentration jump across the interface.  Such a concentration jump leads to a density jump across the interface which generates a potential energy cost to further mixing. On the other hand, heating of the convection zone by the imposed heat flux through the lower boundary causes a temperature jump to form that reduces the magnitude of the density jump induced by mixing.

Thermal diffusion through the convective boundary cools the convection zone and hence increases the density jump. Chemical diffusion has two possible effects: diffusion across the interface removes mass from the convection zone hence reducing the density jump.  But, the mass that is removed is extracted from the upper portion of the convection zone and this mass deficit must be mixed throughout the convection zone.  This mixing acts as an additional cost on the kinetic energy flux and effectively reduces the amount of energy available for entrainment.  The first effect (reducing the density jump) speeds up the convective interface, while the second reduces the speed.  We find that the reduction of the available energy for lifting work is the dominant effect and compositional diffusion acts as a net inhibitor of the entrainment rate.

Recent laboratory experiments by \citet{Dorel2023} have measured the advance of a convective interface that is remarkably fast. They find that the speed of the convective interface grows linearly with time ($dh/dt \propto t$ or $h\propto t^2$). However, we note that the classic scaling law for the depth of the convection zone $h \propto t^{1/2}$ is a direct consequence of three assumptions: (1) the initial density profile is a linear function of height, (2) the heat flux at the boundary is constant, and (3) there is an absence of diffusive restratification above the convective front. The linear stratification ensures that the density jump scales as $\Delta\rho \propto h$ and the associated potential energy barrier as $E \propto h^3$. The constant heat flux results in a kinetic energy flux that scales linearly with the depth of the convection zone, $U_c^3 \propto F_* h$. When the entrainment hypothesis is applied, the resulting scaling is $h^2 dh/dt \propto F_* h$ or $h\propto t^{1/2}$. The assumption regarding restratification is made for the sake of tractability but is usual valid when diffusive effects are negligible.

In the laboratory experiment of \citet{Dorel2023} all three of these classic assumptions are invalid: (1) the initial density profile is Gaussian with the peak at the bottom of the domain, (2) the boundaries have constant temperature not constant heat flux, and (3) chemical diffusion leads to significant restratification above the interface. The first two of these differences should lead to acceleration of the interface compared to the classic result. In the absence of restratification, upward mixing of a Gaussian profile will initially result in a density jump that grows with time.  However, once the convective interface has moved through the core of the Gaussian, the density of solute in the wings is so low that additional mixing does not appreciably change the amount of solute in the convection zone. Therefore, as the convection zone spreads further into the wing, the density jump eventually decreases as the reciprocal of the convection zone's thickness, $\Delta\rho \propto 1/h$. For fixed-flux boundary conditions, this density jump would result in an entrainment equation of the form, $dh/dt \propto F h$, which suggests even more rapid acceleration of the convective boundary, i.e., $h \propto \exp(\Gamma t)$ where $\Gamma$ is a positive constant.

Fixed-temperature boundary conditions lead to further acceleration of the convective interface as the heat flux itself increases as the convection zone thickens. If we assume a Nusselt number that scales with the Rayleigh number as a power law, ${\rm Nu} \sim \Ra^\delta$, one arrrives at a flux that scales with the depth of the convection zone, $F\propto h^{3\delta-1}$. Further, laboratory and numerical experiments suggest that $1/3\le\delta\le1/2$ (see the review by \citet{Plumley2019}).  Thus, for the lower bound, $\delta = 1/3$, the flux is constant as the convection zone deepens, but for any larger value, there is a weak increase in the flux.  For example, at the upper bound, $\delta = 1/2$, the flux scales as the square root of the convection zone thickness, $F\propto h^{1/2}$ leading to even faster acceleration of the convective interface.

In the experiments of \citet{Dorel2023} the combination of the Gaussian profile of the solute concentration and the fixed temperature boundary conditions should result in a convective interface that accelerates at a hyper-exponential rate, we should not be asking why the growth rate in the experiment $h \sim t^2$ is so much faster than the classical result $h \sim t^{1/2}$.  Instead, we should be asking why the growth rate in the experiment was so slow. We suggest that compositional diffusion is the reason.  As we discussed earlier for our numerical simulations, compositional diffusion modifies the speed of the convective interface in several competing ways. By removing solute from the convection zone and adding it to the overlying stable region, diffusion weakens the density jump at the interface, reduces the potential energy barrier, and accelerates the interface.  Conversely, when diffusion removes solute from the top of the convection zone, the resulting mass deficit must be mixed throughout the convection zone and this requires energy. Thus, compositional diffusion reduces the amount of kinetic energy that is available to do further mixing work, thereby slowing the interface speed. Our supposition is that this later mechanism (the reduction of the available kinetic energy for mixing) dominates in the experiment, slowing the interfaces's speed from exponential growth to a power law.  Such a result is self-consistent with the simulations performed here, as this mechanism also dominates in our work.

In our simulations, when the diffusivities are sufficiently large, diffusive spreading ahead of the convective boundary outpaces the boundary’s motion. As a result, the stable layer above the interface can become significantly restratified before the boundary arrives. Further, the restratification allows the stable region to diffusively equilibrate and if the boundary conditions are permissive,  a quasi-steady state can be achieved once a diffusion time has passed.  This is the cause of the convective stalls that occur in our simulations.  These models either have large compositional diffusion (and hence rapid diffusive spread and a slower interface) or they are extremely stable with large buoyancy frequency and hence very slow convective interfaces. In our models the boundary conditions permit quasi-steady state solutions, by allowing purely linear profiles of the heat and compositional fluxes. In the convection zone, the temperature and concentration are essentially uniform, whereas in the overlying stable region, these profiles are parabolic (such that their gradients are uniform). Within the convective boundary the profiles smoothly transition from one functional form to the other.

Based on typical values of the chemical diffusivity in stars and planets, where the Lewis number is enormous, $\mathrm{Le}\sim \mathrm{Pr}^{-1} \sim 10^2$--$10^6$, chemical diffusion is too slow on its own to halt the propagation of a convective boundary. However, if a stall should occur for other reasons (such as photon diffusion in stars undergoing core convection), chemical diffusion could be an important mechanism determining the shape of the convective interface and in modifying the flux of heavy elements across that interface. 
Gas giants are a particularly interesting case as the heat flux that drives convection decreases over time (since the luminosity drops as the planet cools), and the extreme rotation of these objects reduces the kinetic energy flux available for mixing (see \cite{Fuentes2023,Fuentes2024}). Both effects have been shown to reduce the propagation rate of the convective boundary and, in a variety of relevant cases, can lead to a complete stall of the convective zone (see \cite{Hindman2023, Zhang2025}). Due to the slow entrainment rate in such objects, diffusion may play a more active role, slowing the convective interface even further, changing the spatial structure of the convective boundary, and modifying the regime boundaries in parameter space for stalling behavior.  

We have made two key model simplifications which, if relaxed in future work, could significantly improve and inform models of stellar and planetary evolution. The first is the use of the Boussinesq approximation, which assumes nearly incompressible fluids with small density variations. In contrast, real stars and planets span many density scale heights, and this stratification undoubtedly alters the potential energy budget required for mixing. As a result, it would likely change the scaling of the convective zone's growth rate. A second simplification is the use of Cartesian geometry. Real astrophysical objects are spherical, rotate, and exhibit radially varying gravity profiles. Recent simulations of semiconvection in rotating spheres have demonstrated that both the misalignment between gravity and rotation as a function of latitude, and the radial decline in gravity close to the center, can significantly influence the buoyancy flux needed for mixing \cite{Fuentes2025}. These effects also lead to anisotropic mixing, which could, in principle, modify the entrainment rates inferred in earlier studies and affect the overall growth of convective layers. Numerical simulations going beyond Boussinesq and adopting spherical geometry would be of great interest.\\

%%%%%%%%%%%%%%%%%%%%%%%%%%%%%%%%%%%%%%%%%%%%%%%%%%%%%%%%%%%%%%%%%%%%%%%%%
%               Acknowledgements                                        %
%%%%%%%%%%%%%%%%%%%%%%%%%%%%%%%%%%%%%%%%%%%%%%%%%%%%%%%%%%%%%%%%%%%%%%%%%
\begin{acknowledgements}
 B.W.H. and J.R.F. are supported by NASA Solar System Workings grant 80NSSC24K0927. B.W.H. is further supported by NASA grants 80NSSC24K0125 and 80NSSC24K0271. J.R.F. is also supported by the Sherman Fairchild Postdoctoral (Burke) Fellowship and the Presidential Fellowship at Caltech.
\end{acknowledgements}

%%%%%%%%%%%%%%%%%%%%%%%%%%%%%%%%%%%%%%%%%%%%%%%%%%%%%%%%%%%%%%%%%%%%%%%%%
%               Bibliography                                            %
%%%%%%%%%%%%%%%%%%%%%%%%%%%%%%%%%%%%%%%%%%%%%%%%%%%%%%%%%%%%%%%%%%%%%%%%%
\bibliography{references}

\end{document}

%% file: figures.tex
\def\figschematic{%
\begin{SCfigure}
    \centering
    \includegraphics[width=0.7\linewidth]{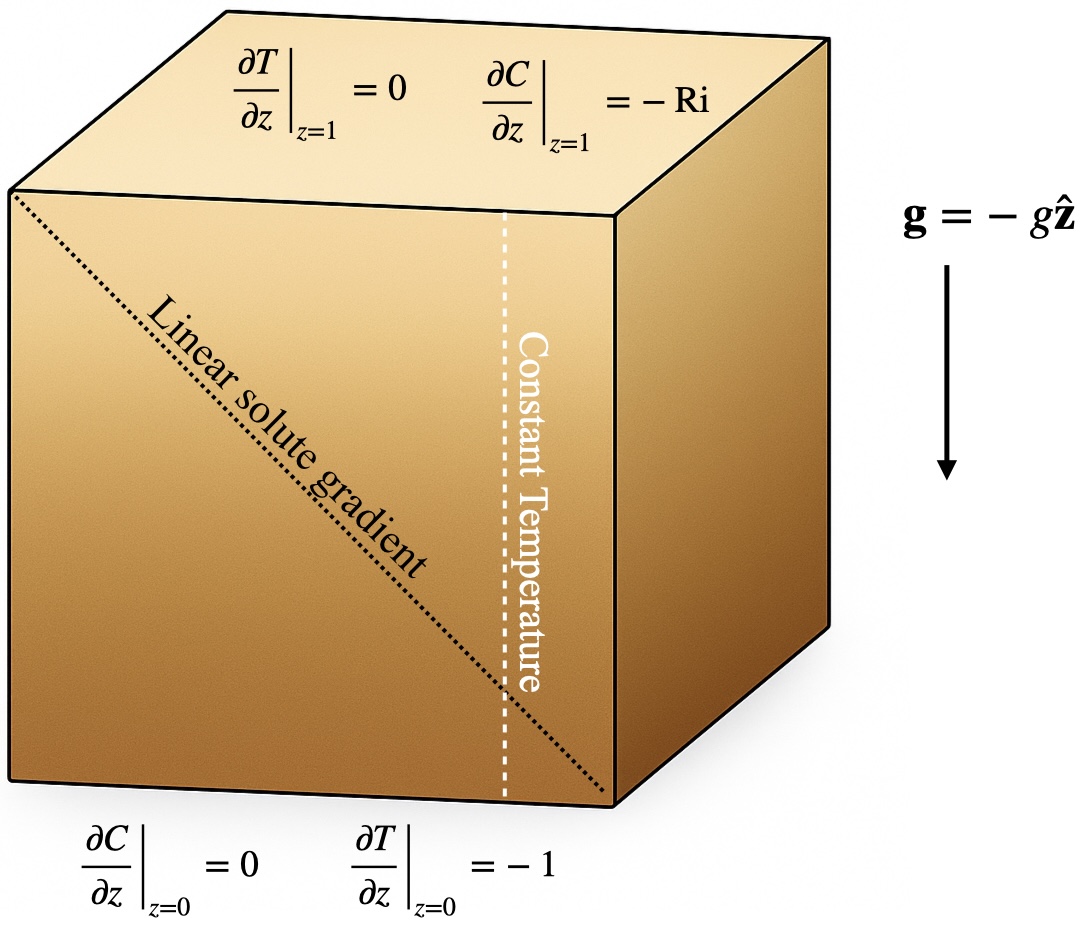}
    \caption{Geometrical set-up of the simulations. A Cartesian box with a stably stratified solute gradient is heated at a constant rate at the bottom boundary. The horizontal boundary conditions are periodic, and the boundary conditions for the velocity field are impenetrable, stressfree bottom and top boundaries.}
    \label{fig:scheme}
\end{SCfigure}
}

\def\figregimes{%
\begin{figure}
    \centering
    \includegraphics[width=0.7\textwidth]{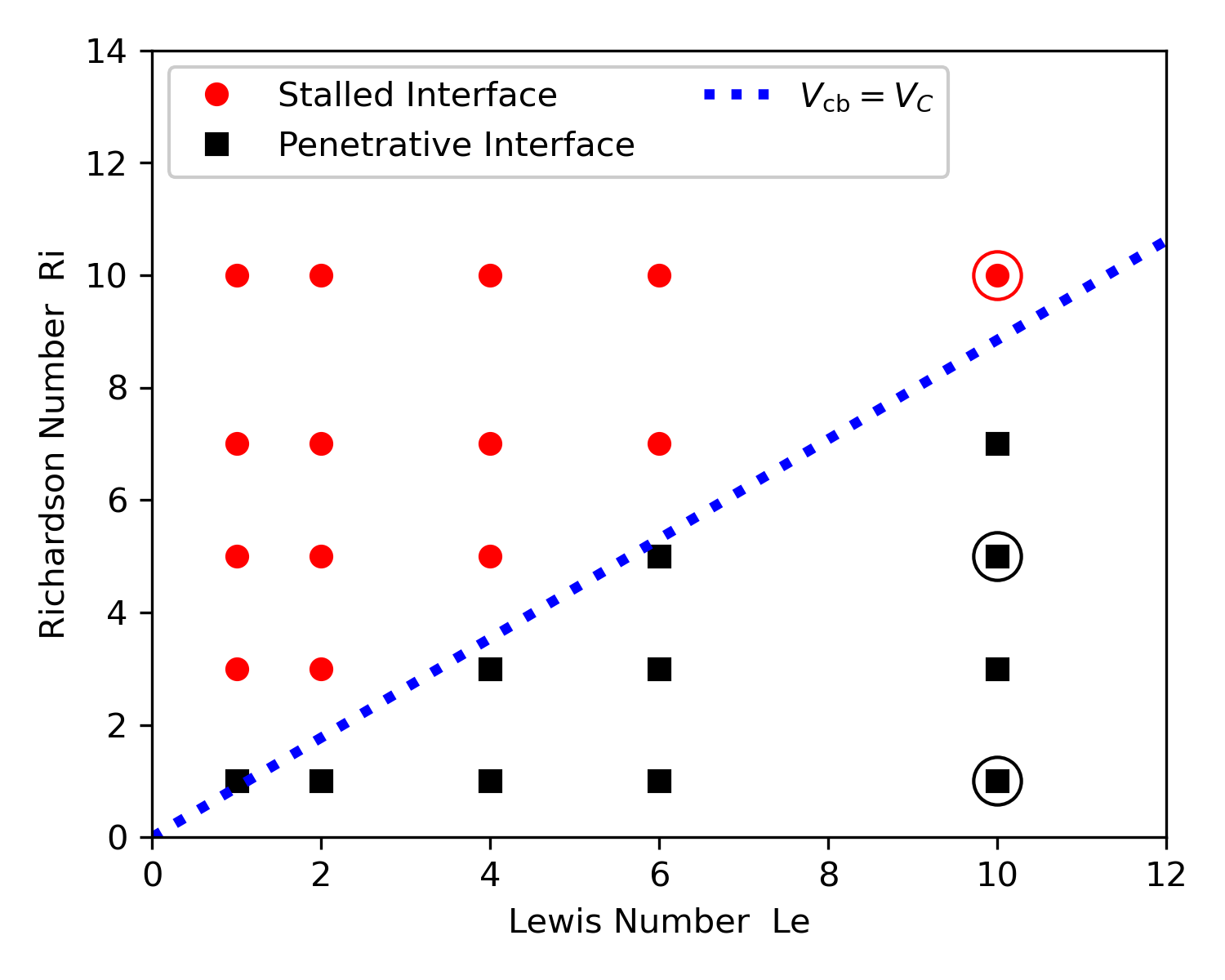}
    \caption{The behavior of the suite of simulations as a function of the Lewis Number, $\Le$, and the Richardson Number, $\Ri$. All of the simulations have the same flux Rayleigh Number, $\Ra= 10^8$, and Prandtl Number, $\Pr = 0.1$. The black squares indicate 2D simulations that exhibit penetrative convection with a rapidly ascending convective boundary that eventually chews through the entire computational domain. Red disks correspond to 2D simulations that eventually reach a diffusive quasi-equilibrium where the convective interface stalls; these models enter a regime of overshooting convection. The large circles correspond to 3D simulations that also exhibit penetrative (black) or stalled (red) convective interfaces. The blue dotted line corresponds to parameter values where the analytically derived speed of the convective interface matches the compositional diffusion speed, see Equation~(\ref{eqn:Vcb=VC}).}
    \label{fig:regimes}   
\end{figure}
}

\def\figfaststills{%
\begin{figure*}
    \centering
    \includegraphics[width=\textwidth]{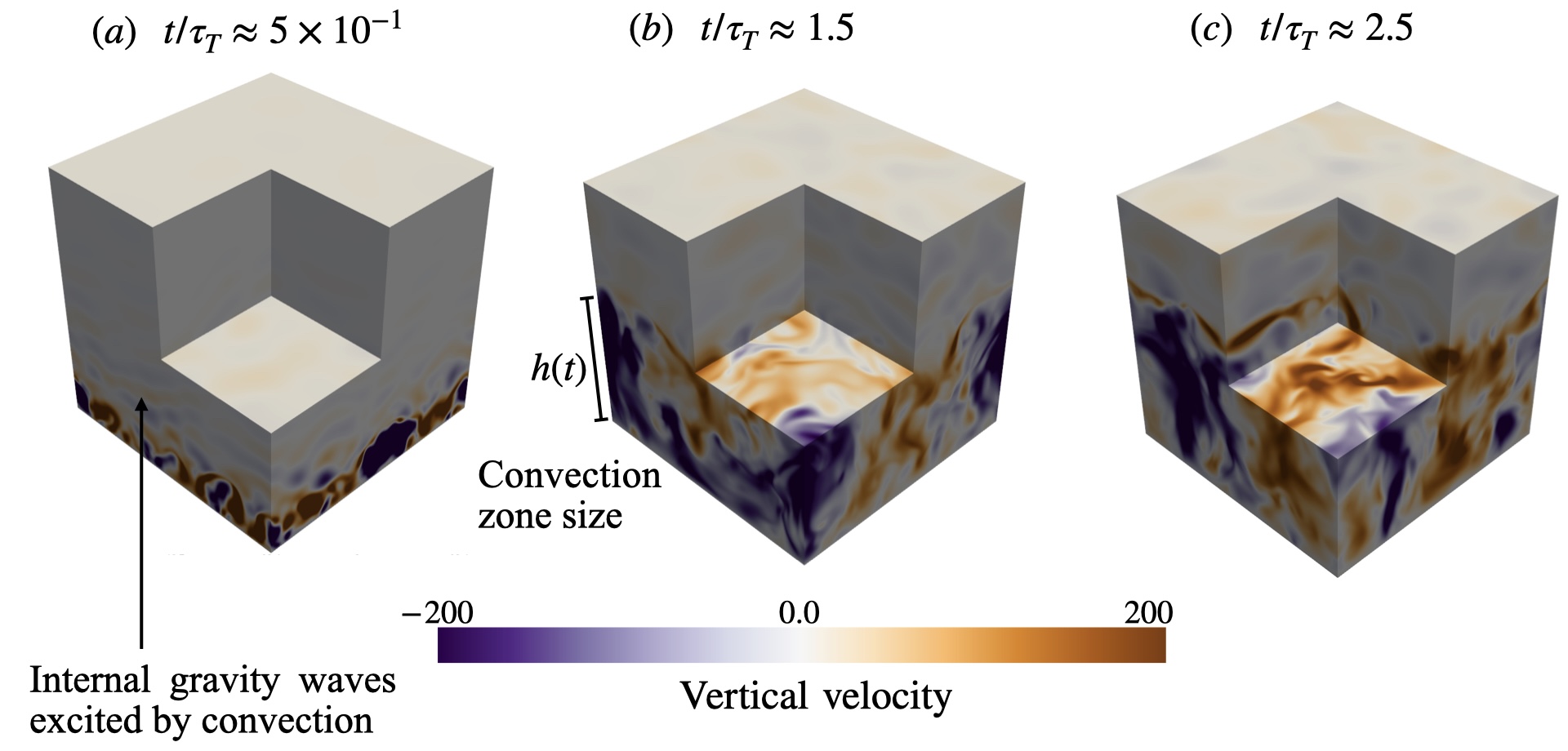}
    \vspace{-0.4in} 
    \caption{Snapshots from a 3D simulation exhibiting penetrative convection. The simulation has a relatively large initial stratification and weak compositional diffusion ($\Ri=5$, $\Le=10$). Each image shows the vertical component of the velocity. Three distinct times have been illustrated: (a) An early time just after the onset of convection.  (b) An intermediate time when the convection zone is fully formed and has entrained roughly half the spatial domain.  (c) A late time when the convection zone has entrained about 75\% of the domain.  In all snapshots, the convection evinces fast upflows and downflows, but even in the overlying stable layer weak flows appear due to the presence of internal gravity waves excited by pummeling of the convective boundary by rising plumes.
    }
    \label{fig:fast_stills}

\end{figure*}
}

\def\figfastprofs{%
\begin{figure*}
    \centering
    \includegraphics[width=\textwidth]{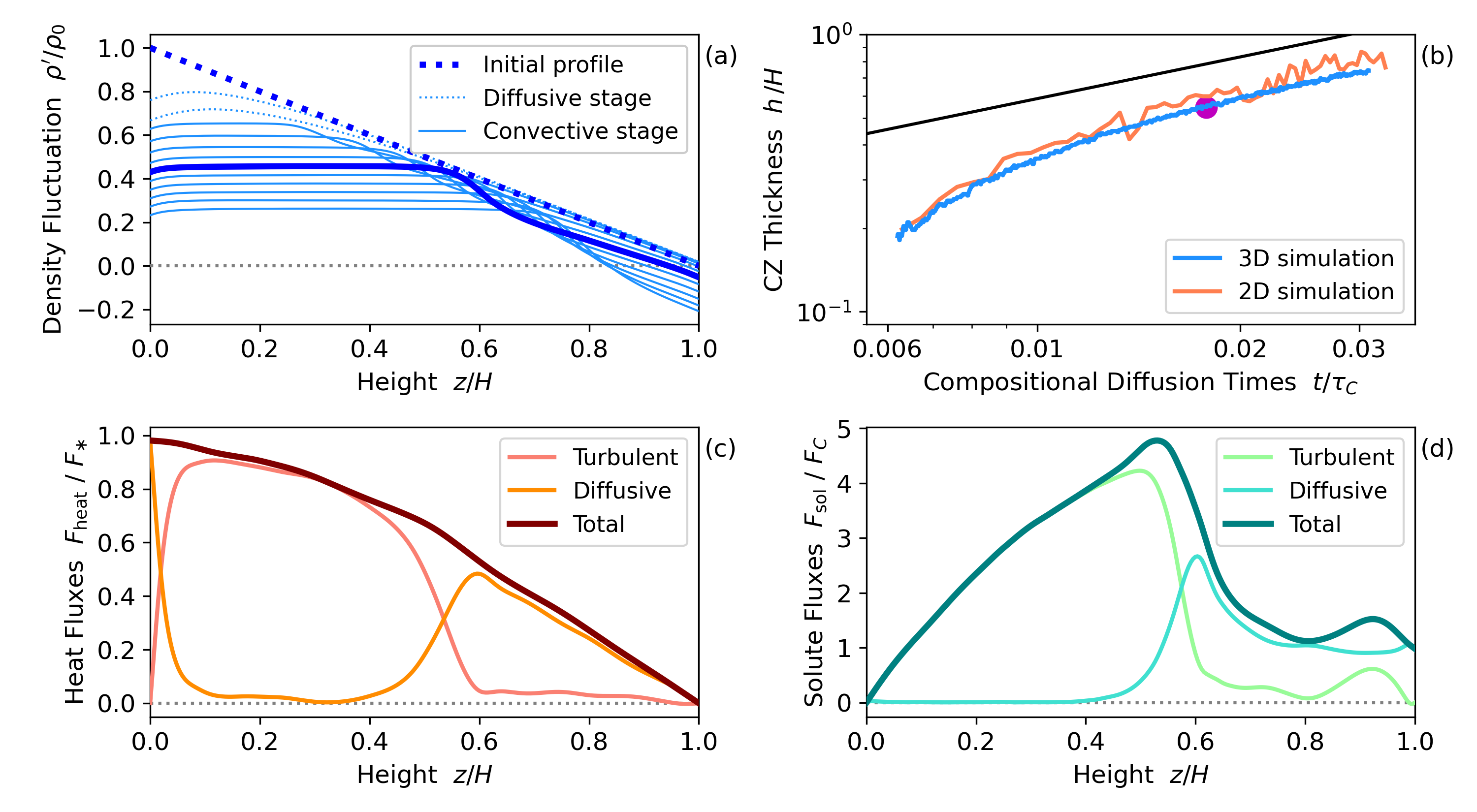}
    \setlength{\abovecaptionskip}{-15pt}

    \caption{Vertical profiles and a time trace from a 3D simulation ($\Ri = 1$, $\Le = 10$) exhibiting penetrative convection. The vertical profiles have been measured by averaging the illustrated field horizontally and temporally. (a) Vertical profiles of density fluctuations induced by composition and temperature $\rho^\prime/\rho_0 = \beta C-\alpha T$. The dotted curves correspond to early times during the diffusive stage before the onset of convection, whereas the solid curves indicate later times during the convective stage. The thick dotted curve corresponds to the initial stratification and the thick solid curve is for a single time which we have emphasized to make the profile shapes readily apparent. (b) The thickness of the convection zone as a function of time for both the 2D and 3d models. Only times after the convection zone is formed are shown. The time emphasized in panel a is indicated with the violet circle. The black reference line indicates a power law $h\propto t^{1/2}$. (c) Profiles of the vertical heat flux normalized by the heat flux imposed at the lower boundary, $F_*$. The diffusive and turbulent fluxes are shown separately, as well as added together. (d) Profiles of the vertical composition flux normalized by the flux imposed at the upper surface, $F_C=\kappa_C (\beta g)^{-1} N_0^2$. Both the thermal and compositional fluxes are extracted from the simulation at the time indicated by the violet circle in panel b.
    }
    \label{fig:fast_profs}

\end{figure*}
}

\def\figspeed{%
\begin{figure}
    \centering
    \includegraphics[width=\textwidth]{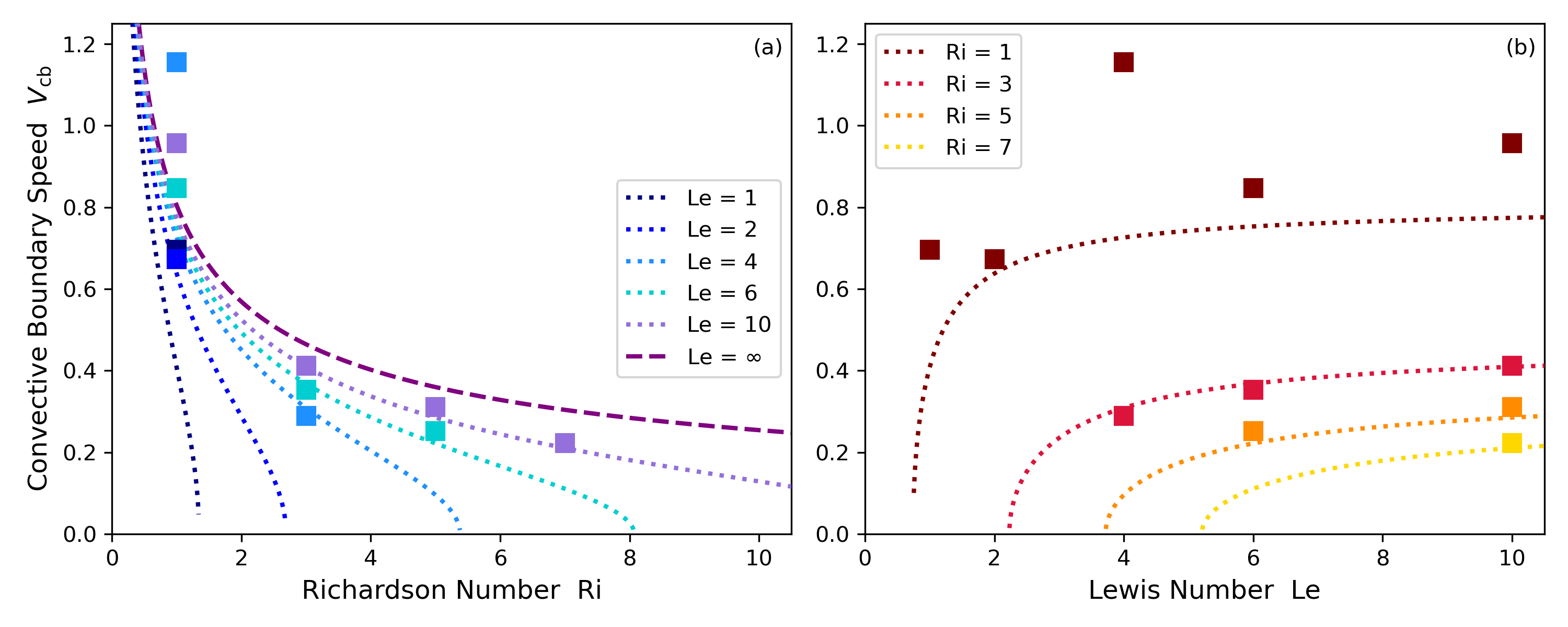}
    \caption{Nondimensional convective boundary speed from the simulations (squares) and from theory (curves) as a function of the Richardson and Lewis numbers. As indicated in the caption, the colors indicate either the Lewis number or Richardson number depending on the panel. The convective boundary speed was obtained from those simulations undergoing penetrative convection by fitting the height of the convective boundary with a power law in time, $h = q^{-1} V_{\rm cb} \, (\kappa_T t)^q$. The theoretical curves come from Equation~(\ref{eqn:Vi}), which is applicable in the weak diffusion limit.}
    \label{fig:Vcb}
\end{figure}
}

\def\figstalledprofs{%
\begin{figure*}
    \centering
    \includegraphics[width=\textwidth]{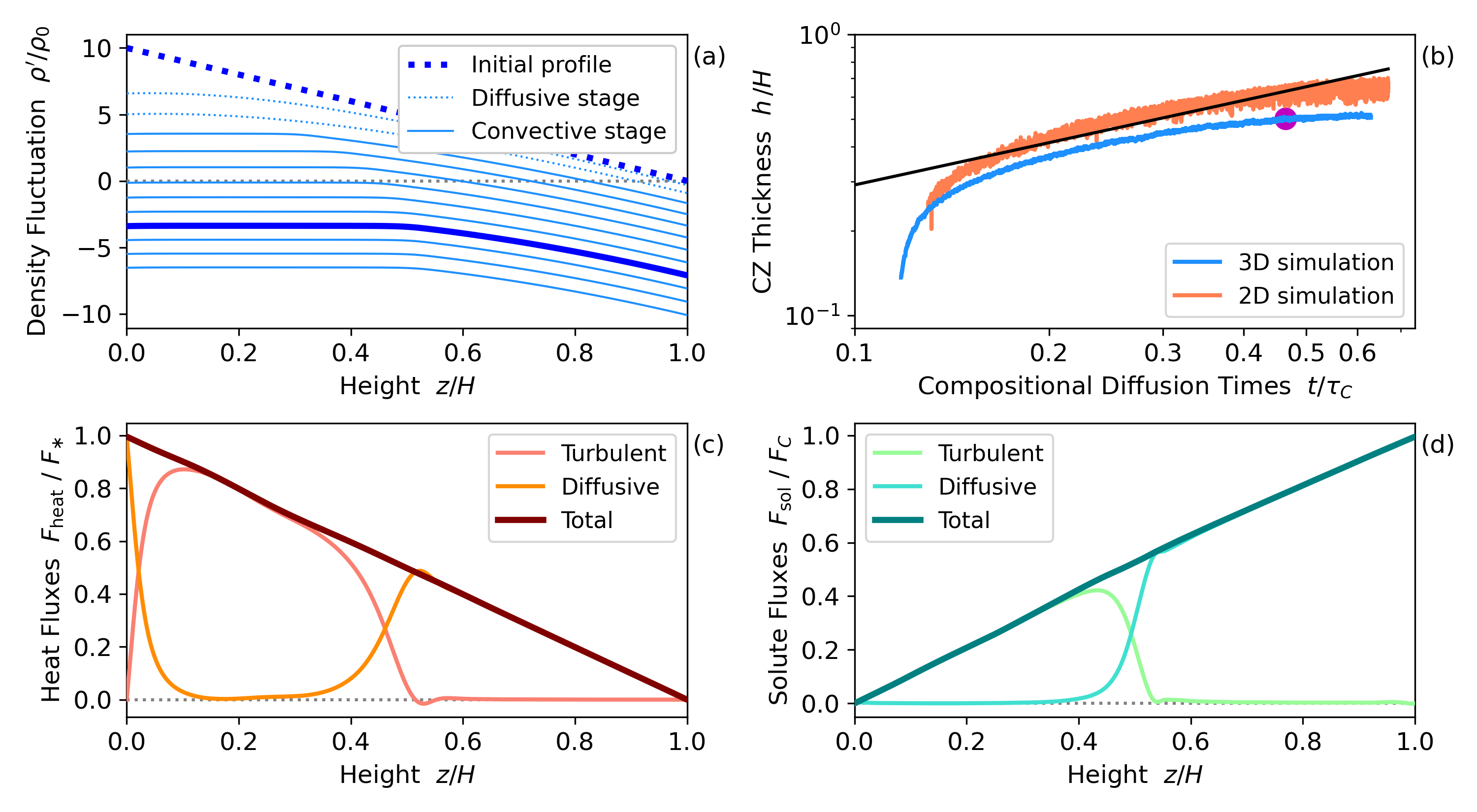}
    \setlength{\abovecaptionskip}{-15pt}
    \caption{Vertical profiles and time traces from a 3D simulation ($\Ri = 10$, $\Le = 10$) exhibiting a stalled convective boundary. The curves and colors have the same meaning as in Figure~\ref{fig:fast_profs}. (a) Profiles of density fluctuations.  (b) The height of the convective boundary as a function of time for both the 3D and 2D models with $\Ri = 10$ and $\Le = 10$. (c) Profiles of the normalized vertical heat flux. (d) Profiles of the normalized vertical composition flux. Both of the fluxes are extracted from the simulation at the time indicated by the violet circle in panel b.
    }
    \label{fig:stalled_profs}

\end{figure*}
}

\def\figstalltime{%
\begin{SCfigure}
    \centering
    \includegraphics[width=0.55\textwidth]{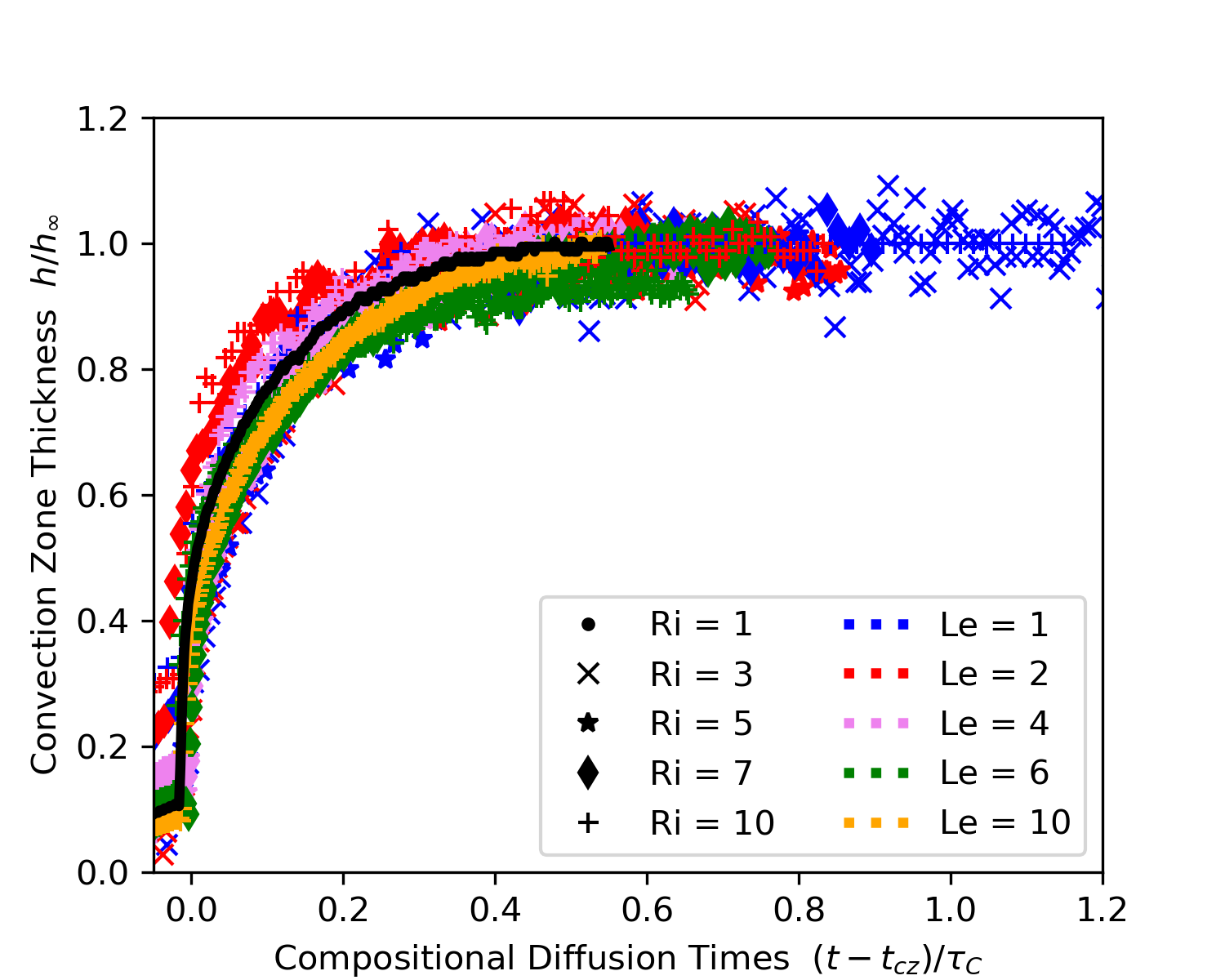}
    \caption{Convection zone thickness for all simulations that exhibit a stalled convective boundary. For the 2D simulations, the colors indicate the Lewis number and the symbols the Richardson number. The black curve corresponds to the 3D simulation with $\Ri = 10$ and $\Le=10$. The height has been normalized by the height achieved for long times to demonstrate that all simulations with a stalled convective boundary have similar temporal behavior. Time is measured in compositional diffusion times elapsed since the formation of the convection zone at $t=t_{cz}$.}
    \label{fig:stall_time}
\end{SCfigure}
}

%% file: packages.tex
\usepackage{ulem}
\usepackage[varg]{txfonts}
\usepackage{bm}
\usepackage{float}
\usepackage{setspace}
\usepackage{xcolor}
\usepackage{sidecap}
\sidecaptionvpos{figure}{t}

\usepackage[left]{lineno}

\newcommand{\cb}{{\rm cb}}

\newcommand{\grad}{\bm{\nabla}}

\newcommand{\Ri}{\mathrm{Ri}}
\newcommand{\Ra}{\mathrm{Ra}}
\newcommand{\Le}{\mathrm{Le}}

\renewcommand{\bm}[1]{{\mbox{{\boldmath$#1$}}}}	% bold in mathmode

\usepackage[caption=false]{subfig}
\PassOptionsToPackage{hyphens}{url}
\usepackage[bookmarks=true,         % show bookmarks bar?/ Changed March 22, 2019 for
     pdfnewwindow=true,      % links in new window
     colorlinks=true,    % false: boxed links; true: colored links
     linkcolor=blue,     % color of internal links
     citecolor=blue,     % color of links to bibliography
     filecolor=blue,  % color of file links
     urlcolor=blue,      % color of external links
     final=true,
 ]{hyperref}
\usepackage{color,epsfig,amsmath,amssymb,fancyhdr} %\usepackage[T1]{fontenc}

\usepackage{footnote}
\newcommand\footnoteref[1]{\protected@xdef\@thefnmark{\ref{#1}}\@footnotemark}
\makeatother
\usepackage{etoolbox}
\usepackage{footmisc}
\makeatletter
\patchcmd{\frontmatter@abstract@produce}
  {\vskip200\p@\@plus1fil
   \penalty-200\relax
   \vskip-200\p@\@plus-1fil}
  {}
  {}
  {}
\makeatother

\usepackage{booktabs}